\documentclass[%
prl,final,superscriptaddress,twocolumn, floatfix
]{revtex4-2}

\usepackage{graphicx}
\usepackage{dcolumn}
\usepackage{bm}
\usepackage{siunitx}
\usepackage{tabularx}
\usepackage{latexsym}
\usepackage{changes}
\usepackage{esvect}
\usepackage[colorlinks = true,
            linkcolor = blue,
            urlcolor  = blue,
            citecolor = blue,
            anchorcolor = blue]{hyperref}
\usepackage{mathtools}
\usepackage{xcolor}
\usepackage{soul}
\usepackage{amssymb}
\usepackage{bm}
\usepackage{color}
\usepackage{braket}
\usepackage{dsfont}
\usepackage{wasysym}
\usepackage{xfrac}
\usepackage{braket}
\usepackage{textgreek}
\usepackage{ulem}
\usepackage{cleveref}
\usepackage{adjustbox}

\usepackage{makecell}

\begin{document}
\setcounter{figure}{0}
\renewcommand{\thetable}{\arabic{table}}

\title{Effects of Laser-Annealing on Fixed-Frequency Superconducting Qubits}

\author{Hyunseong Kim}
\email[Email:]{hyunkim@berkeley.edu}
\affiliation{Department of Physics, University of California, Berkeley, CA 94720, USA}

\author{Christian J{\"u}nger}
\affiliation{Computational Research Division, Lawrence Berkeley National Laboratory, Berkeley, California 94720, USA}

\author{Alexis Morvan}
\affiliation{Computational Research Division, Lawrence Berkeley National Laboratory, Berkeley, California 94720, USA}

\author{Edward S. Barnard}
\affiliation{Molecular Foundry Division, Lawrence Berkeley National Laboratory, Berkeley, California 94720, USA}

\author{\\ William P. Livingston}
\affiliation{Department of Physics, University of California, Berkeley, CA 94720, USA}

\author{M. Virginia P. Altoé}
\affiliation{Molecular Foundry Division, Lawrence Berkeley National Laboratory, Berkeley, California 94720, USA}

\author{Yosep Kim}
\affiliation{Computational Research Division, Lawrence Berkeley National Laboratory, Berkeley, California 94720, USA}

\author{Chengyu Song}
\affiliation{Molecular Foundry Division, Lawrence Berkeley National Laboratory, Berkeley, California 94720, USA}

\author{Larry Chen}
\affiliation{Department of Physics, University of California, Berkeley, CA 94720, USA}

\author{John Mark Kreikebaum}
\affiliation{Department of Physics, University of California, Berkeley, CA 94720, USA}
\affiliation{Materials Science Division, Lawrence Berkeley National Laboratory, Berkeley, California 94720, USA}

\author{D. Frank Ogletree}
\affiliation{Molecular Foundry Division, Lawrence Berkeley National Laboratory, Berkeley, California 94720, USA}

\author{David I. Santiago}
\affiliation{Department of Physics, University of California, Berkeley, CA 94720, USA}
\affiliation{Computational Research Division, Lawrence Berkeley National Laboratory, Berkeley, California 94720, USA}

\author{Irfan Siddiqi}
\affiliation{Department of Physics, University of California, Berkeley, CA 94720, USA}
\affiliation{Computational Research Division, Lawrence Berkeley National Laboratory, Berkeley, California 94720, USA}
\affiliation{Materials Science Division, Lawrence Berkeley National Laboratory, Berkeley, California 94720, USA}

\date{\today}
\begin{abstract}
As superconducting quantum processors increase in complexity, techniques to overcome constraints on frequency crowding are needed. The recently developed method of laser-annealing provides an effective post-fabrication method to adjust the frequency of superconducting qubits. Here, we present an automated laser-annealing apparatus based on conventional microscopy components and demonstrate preservation of highly coherent transmons. In one case, we observe a two-fold increase in coherence after laser-annealing and perform noise spectroscopy on this qubit to investigate the change in defect features, in particular two-level system defects. Finally, we present a local heating model as well as demonstrate aging stability for laser-annealing on the wafer scale. Our work constitutes an important first step towards both understanding the underlying physical mechanism and scaling up laser-annealing of superconducting qubits.
\end{abstract}
\maketitle

\section{Introduction}
\label{sec::sec1}
Superconducting quantum processors are a promising platform for realizing large-scale universal quantum computation \cite{Arute_2019}. In comparison to other physical platforms \cite{Cirac_1995, Loss_1998, Imamog_lu_1999, Hanson_2006, Knill_2001}, superconducting quantum processors are lithographically configurable, which allows a rich variety of qubit structures and feasible scalability, currently up to  $\sim$\,$ 100$ qubits \cite{Krantz_2019, Ball_2021}. Superconducting qubits require the Josephson junction (JJ), a nonlinear inductive element composed of two superconductors with a tunneling barrier in between \cite{Josephson_1962, Josephson_1964}. A capacitively-shunted JJ forms the transmon qubit, a widely utilized superconducting qubit with advantages ranging from high coherence to simple coupling and readout \cite{Koch_2007, Wang_2022}. As quantum processors scale up further, precise fabrication of the JJ is required to avoid qubit frequency allocation problems that can lead to frequency collisions or slow entangling gates \cite{Morvan_2021, Brink_2018, Nguyen_2022}. However, the dispersion of state of the art JJ fabrication methods, currently $\sim1\%$ for a $\SI{1}{\centi\meter}^2$ chip, does not suffice the frequency constraints for a fixed-frequency multiqubit processor with even a few tens of qubits \cite{Kreikebaum_2020}. One effective technique to circumvent this problem is post-fabrication laser-annealing, in which a laser beam is applied to the JJ in order to tune the qubit frequency \cite{Hertzberg_2021, Zhang_2022}. In this work, we build upon this technique and present a laser-annealing apparatus with conventional confocal microscopy components, allowing integration into various qubit preparation processes. We demonstrate that high coherence of fixed-frequency transmon qubits is maintained after frequency shifting by laser-annealing. In one case, we observe a two-fold increase in coherence after laser-annealing. We perform noise spectroscopy to investigate this increase in coherence by comparing the change in two level system defect features and suggest that the increase in coherence may be correlated to the
decrease in spectrally neighboring TLS’s after laser-annealing. Lastly, we successfully perform laser-annealing on the wafer scale with varying parameters and explain the results by using a local heating model, as well as demonstrate the stability of laser-annealing with respect to JJ aging.\par
%
%
\begin{figure}[h!]
\includegraphics[width=\columnwidth]{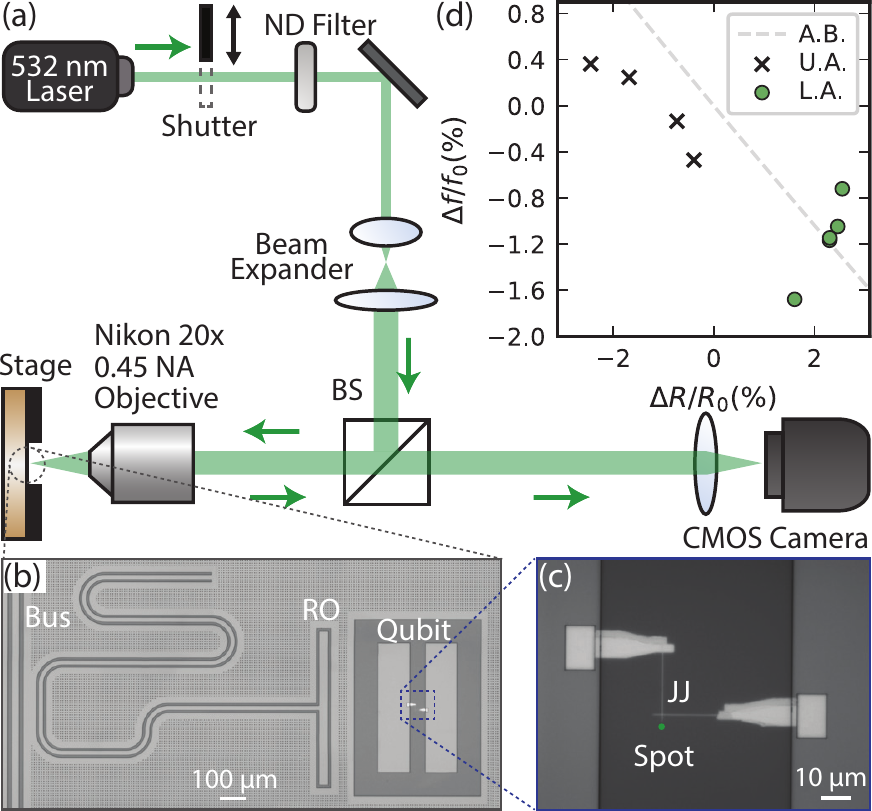}
\caption{(a) Schematic of laser-annealing apparatus. The continuous wave, \SI{532}{\nano \meter}, \SI{40}{\milli \watt} diode-pumped solid state laser beam follows the optical path shown by the green arrows and is focused down to a beam waist of \SI{0.81}{\micro \meter}. Abbreviated optical components are labeled as follows:  Neutral Density (ND) filter, Beam Splitter (BS), and Numerical Aperture (NA). (b) Optical image of fixed-frequency transmon qubit, consisting of Al/Al-O\textsubscript{x}/Al Josephson junction (JJ), shunted by niobium (Nb) planar capacitor. The qubit is capacitively coupled to a Nb quarter-wavelength coplanar waveguide resonator (RO) and read out through the bus. (c) Optical magnification of JJ. Laser spot depicted by green circle and drawn to scale. (d) Predicted (gray dashed line derived from Ambegaokar-Baratoff (A.B.) formula) and measured qubit frequency shift ($\Delta f/f_0$) with respect to change in room temperature resistance ($\Delta R/R_0$) due to laser-annealing (L.A.). Unannealed samples are labeled U.A.}
	\label{fig:fig1}
\end{figure}
\section{Laser-Annealing Apparatus And Characterization}
\label{sec::sec2}
In order to facilitate integration into various qubit preparation processes, we present a laser-annealing apparatus based on confocal microscopy (schematically shown in Fig.~\ref{fig:fig1}(a)). A continuous, collimated laser beam (\SI{532}{\nano \meter}) follows the optical path depicted by the green arrows and first passes through a shutter that serves as a switch and then a neutral density filter that adjusts the beam power. The beam is then expanded and focused onto the sample stage in order to minimize the beam spot. The motorized stage automatically positions the sample on the focal plane using image detection algorithms. Imaging is performed with a CMOS camera and a white light source (not shown). This automation enables laser-annealing of both individual chips as well as \SI{100}{\milli\meter} wafers with 3000 JJs.\par
The sample under investigation is composed of four independent transmon qubits each at  different frequencies and capacitively coupled to a coplanar waveguide resonator (RO). A representative qubit-RO pair is shown in Fig.~\ref{fig:fig1}(b). The qubit consists of a niobium (Nb) coplanar capacitor that shunts an Al/Al-O\textsubscript{x}/Al JJ (Fig.~\ref{fig:fig1}(c)). The JJs are shadow evaporated using the Manhattan style technique with typical areas of  $\sim\SI{0.1}{\micro\meter}^2$. The JJ areas are varied to differ the frequencies of the four qubits. To isolate the effects of laser-annealing, we minimize resonator-induced decay and drive the qubit through the readout bus (RO) \cite{Krantz_2019}. Further details regarding the setup and fabrication are given in the Supplemental Materials (S.M.).\par
We apply laser-annealing to the JJ's of five transmon qubits and investigate the response in normal state resistance $R_\textrm{N}$ at room temperature and the resulting shift in qubit frequency $f_\textrm{Q}$ at $\sim$\,$\SI{20}{\milli \kelvin}$. We measure $R_\textrm{N}$ using a lock-in voltage probe with the probing needles electrically contacting the Nb capacitors. From $R_\textrm{N}$, we utilize the Ambegaokar-Baratoff formula to calculate the critical current $I_\textrm{C}$, nominally around \SI{35}{\nano\ampere} \cite{Ambegaokar_1963}. Applying the transmon-regime approximation with $I_\textrm{C}$, the predicted qubit frequency in the superconducting state is given by:
\begin{equation}
    {hf_\textrm{Q} = \sqrt{(h\Delta_\textrm{Al} E_\textrm{C})/(e^2R_\textrm{N})}-E_\textrm{C}}
\label{eq::eq1}
\end{equation}
where $h$ is Planck's constant, $e$ the electron charge, \textrm{$\Delta\textsubscript{Al}$} the Al superconducting gap (=\SI{170}{\micro \electronvolt}), and $E_\textrm{C}$ the charging energy of the transmon ($E_\textrm{C}/h\sim\SI{275}{\mega\hertz}$) \cite{Krantz_2019}. We expect laser-annealing to increase $R_\textrm{N}$ and resultantly shift down $f_\textrm{Q}$, while \textrm{$\Delta\textsubscript{Al}$} and $E_\textrm{C}$ remain constant.\par 
The normalized change in qubit frequency ($\Delta f/f_0$) is plotted as a function of change in normal state resistance ($\Delta R/R_0$) in Fig.~\ref{fig:fig1}(d), where $f_0$ and $R_0$ are the initial frequency and resistance. The prediction (gray dashed line) is given by: $\Delta f^{\textrm{P}}_{\textrm{Q}}/{{f}^{\textrm{P}}_{\textrm{Q,0}}}=-(1/1.9)\Delta R/R_0$. This is derived from Eq.~\ref{eq::eq1} using $\Delta R/R_0 \ll 1$ and applying a 5\% correction to the slope due to $E_\textrm{C}$. Both unannealed (black cross) and laser-annealed (green circles) qubits follow the trend of the prediction, with a controlled frequency downshift for the laser-annealed qubits. The frequency shift of the four unannealed qubits, due to air reexposure during the laser-annealing step and frequency fluctuations across cryostat cooldowns, average to zero with a variation of $\pm0.3\%$ \cite{McRae_2021}. The resistance drift of the unannealed qubits as well as the discrepancies between prediction and measurement may be due to differences between the expected and actual values of {$\Delta_\textrm{Al}$} and $E_\textrm{C}$, as well as  electrical contact variations across multiple resistance probings.\par
\section{High Coherence and two-level System Spectroscopy}
\label{sec::sec3}
\begin{figure}[h]
\includegraphics[width=\columnwidth]{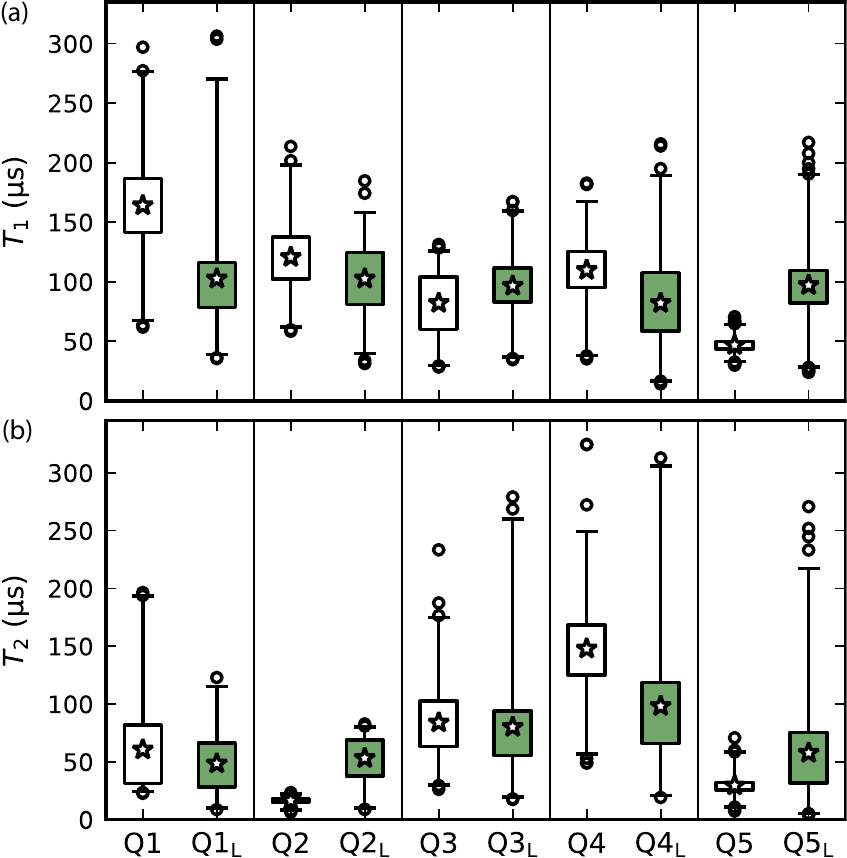}
\caption{Comparison of (a) $T_\textrm{1}$ and (b) $T_\textrm{2}$ coherence times of four qubits  before (Q1-Q5) and after laser-annealing (Q1\textsubscript{L}-Q5\textsubscript{L}). Each boxplot consists of the mean (star), interquartile range (box boundaries), three standard deviation range (caps), and outliers (white circles) of the \SI{17}{hour} acquisitions of each qubit. The $T_\textrm{2}$ measurements of Q2 yielded poor results and large fitting errors, leading to low statistics. Q5 exhibits a two-fold increase in coherence and is used for TLS spectroscopy.}
\label{fig:fig2}
\end{figure}

We have so far demonstrated tunability of qubit frequency using laser-annealing. We now evaluate the quality of laser-annealed qubits. In particular, the qubit relaxation ($T_\textrm{1}$) and phase coherence ($T_\textrm{2}$) times are measured since these metrics are highly sensitive to degradation in materials quality \cite{Lisenfeld_2019}. To study the statistical features, we acquire the $T_\textrm{1}$ and $T_\textrm{2}$ of the transmons for $\sim$\,$\SI{17}{hours}$ before and after laser-annealing. The $T_\textrm{1}$ acquisitions are shown in Fig.~\ref{fig:fig2}(a). Apart from Q5, the coherence medians of laser-annealed qubits Q1\textsubscript{L}-Q4\textsubscript{L} (stars in green boxes) lie within three standard deviations of the medians of Q1-Q4 (caps of white boxes), indicating no statistically significant differences. Similarly, the $T_\textrm{2}$ coherence times do not exhibit any statistically significant differences after laser-annealing (see Fig.~\ref{fig:fig2}(b)). On average, the $T_\textrm{1}$ of Q1\textsubscript{L}-Q4\textsubscript{L} meet the current standards for high coherence times of $\sim$\,$\SI{100}{\micro \second}$ \cite{Siddiqi_2021}. These results verify that our setup successfully performs controlled frequency shifts while preserving high qubit coherence.\par
%
%
In the case of Q5, we observe a statistically significant increase in both $T_1$ (\SI{46.5}{\micro\second} to \SI{95.0}{\micro\second}) and $T_2$ (\SI{29.0}{\micro\second} to \SI{49.8}{\micro\second}) coherence after laser-annealing. In contrast to Q1-Q4, Q5 has an additional CPW to drive the qubit, enabling noise spectroscopy that can help investigate the increase in coherence. Several different noise sources can hinder coherence, such as dielectric loss, quasi-particle tunneling, and cosmic radiation \cite{Siddiqi_2021, Wang_2014, Veps_l_inen_2020}. In particular, losses due to dielectrics at the metal-air, metal-substrate, and substrate-air interfaces of superconducting qubits can transversely couple and potentially induce energy relaxations \cite{Lisenfeld_2019}. These dielectric losses can be modeled as two-level systems (TLS) with transition frequencies $f_\textrm{TLS}$ and coupling $g$ to the qubit. The qubit relaxation rate $\Gamma\textsubscript{1}=1/{T}_\textrm{1}$ increases the closer $f_\textrm{Q}$ is to $f_\textrm{TLS}$. In particular, \textGamma\textsubscript{1} follows a Lorentzian profile with respect to qubit-TLS detuning $\Delta=f_\textrm{Q}-f_\textrm{TLS}$: ${\Gamma_1 = (2\Gamma g^2)/(\Gamma^2+\Delta^2)+\Gamma_\textrm{1,Q}}$ where $\Gamma$ is the sum of TLS and qubit energy relaxation and dephasing rates and $\Gamma_\textrm{1,Q}$ the frequency-independent qubit energy relaxation rate \cite{Lisenfeld_2019, Barends_2013}. Hence a spectral and temporal sweep of qubit $T_1$, or TLS spectroscopy, can probe the noise environment of a transmon qubit \cite{Klimov_2018, Carroll_2021, B_janin_2021}. We perform TLS spectroscopy on Q5 before and after laser-annealing in order to detect changes in TLS features. We do so by AC Stark shifting the qubit using an off-resonant tone of frequency $f_\textrm{Q}\pm\SI{80}{\mega \hertz}$ \cite{Carroll_2021}. The frequency shift is proportional to the square of the tone amplitude, hence the drive CPW is required to deliver higher power in comparison to driving through the RO \cite{Schneider_2018}. With this configuration, we are able to reliably shift $f_\textrm{Q}$ by $\pm\SI{33}{\mega \hertz}$, measured by a Ramsey sequence. For fast acquisition, we measure the average excited state population ${P}_{\ket{1}}$ around $T_\textrm{1}$.\par 
%
\begin{figure}[t]
\includegraphics[width=\columnwidth]{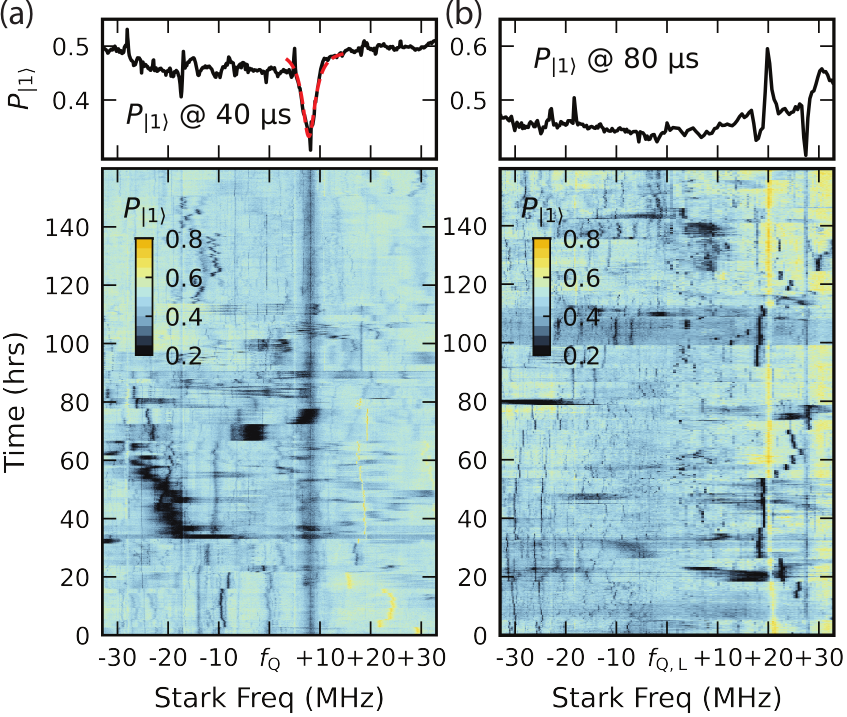}
\caption{(a) Spectral and temporal profile of Q5. Colorbar is scale for excited state population $P$\textsubscript{$\ket{1}$}. Here, $P$\textsubscript{$\ket{1}$} is measured at \SI{40}{\micro\second}. Time-averaged spectral profile is shown on top with fit using Lorentzian around the constant TLS. From the fit, $f_\textrm{TLS}=f_\textrm{Q}+\SI{7.81}{\mega\hertz}$. (b) Spectral and temporal profile of laser-annealed qubit (Q5\textsubscript{L}). Here, $P$\textsubscript{$\ket{1}$} is measured at \SI{80}{\micro\second}. $f_\textrm{Q,L}=f_\textrm{Q}-\SI{94}{\mega\hertz}$. Time-averaged spectral profile on top. No persistent defects are visible.}
\label{fig:fig3}
\end{figure}
TLS spectroscopy of \SI{160}{hours} is shown both before (Fig.~\ref{fig:fig3}(a)) and after (Fig.~\ref{fig:fig3}(b)) laser-annealing. In Fig.~\ref{fig:fig3}(a), one consistent and several fluctuating (dark areas) TLS features are observed close to the initial qubit frequency $f_\textrm{Q}$. Fitting to the Lortenztian, we find the consistent TLS is coupled to the qubit with $g=\SI{76}{\kilo \hertz}$ and lies \SI{7.81}{\mega \hertz} away from $f_\textrm{Q}$, which is more than three linewidths away from $f_\textrm{TLS}$. The low coupling and large spectral distance make it unlikely for this single TLS to solely limit the qubit coherence. If the dominant decoherence channel originates from TLS, it may be due to multiple TLS's that are weakly coupled to the qubit over a wide frequency range around $f_\textrm{Q}$. In contrast, we observe reduced TLS features in the spectral vicinity of the qubit in Fig.~\ref{fig:fig3}(b) after $f_\textrm{Q}$ is downshifted by \SI{94}{\mega\hertz}. We repeat TLS spectroscopy for two more cooldowns to evaluate the spectral features of the qubit due to thermal cycling, and do not observe significant differences with respect to Fig.~\ref{fig:fig3} (see Fig.~\ref{fig:supp_fig1}) \cite{Burnett_2019, de_Graaf_2020}. We suggest that the increase in coherence may be correlated to the decrease in spectrally neighboring TLS's after laser-annealing. Additional studies with a wider spectral range are needed to investigate this. However, our observation opens possibilities to healing a defective qubit on a multiqubit processor using laser-annealing and TLS spectroscopy under the condition that the TLS features are consistent across cooldowns.\par
\section{Laser-Annealing Mechanism}
\label{sec::sec4}
%
%
%
\begin{figure*}[t]
\includegraphics[width=2\columnwidth]{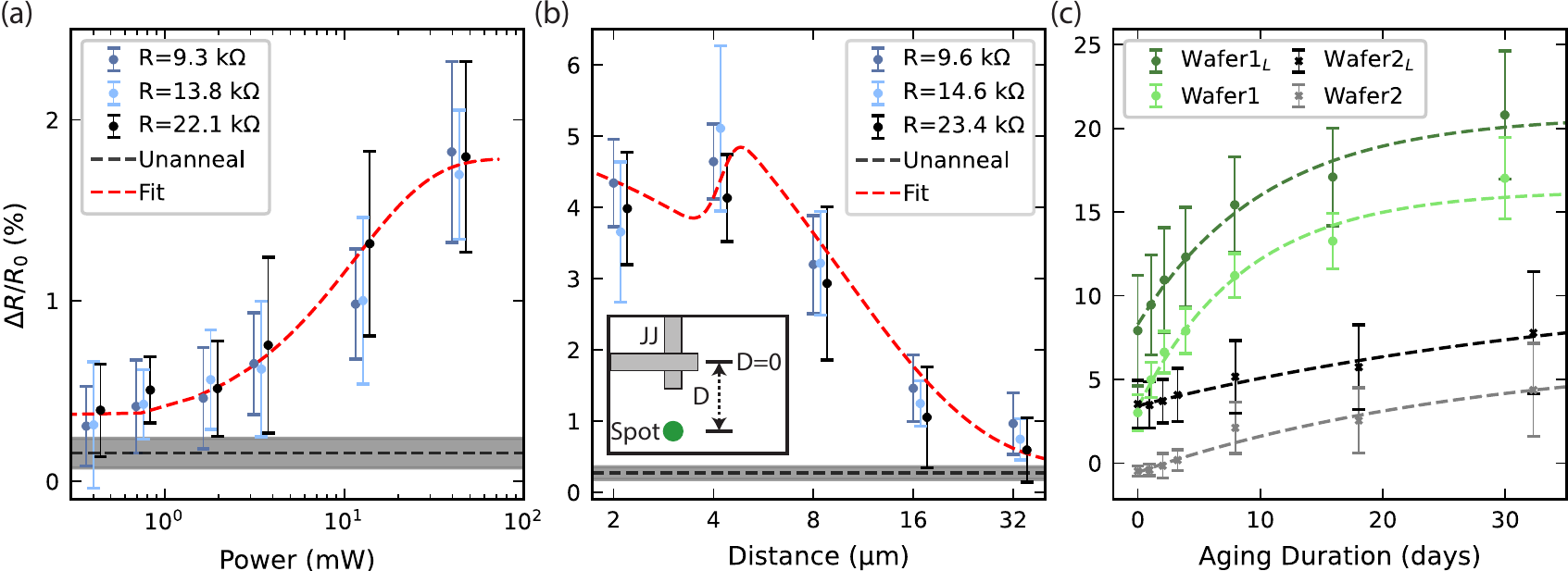}
\caption{Wafer scale characterization of laser-annealing. Different wafers are utilized for each subfigure. (a) Percent JJ resistance change vs laser power. Fit using exponential plateau function of simulated temperature at junction. Each point is the average of $\sim20$ JJs, with the standard deviation given by the errorbars. Unannealed JJs shown by the gray dashed line, with standard deviation shown by the gray shaded region. (b) Percent JJ resistance change vs laser spot displacement from JJ. Maximum resistance change when the spot is displaced \SI{4}{\micro \meter} from the junction. Fit using product of heat transfer and absorbed power. (Inset) Schematic of experiment. Distance ($D$) is measured from below the center of the JJ. (c) Stability of laser-annealing with respect to JJ aging in atmosphere. Each dashed line is data fit to exponentially plateauing function. Wafers 1 and 2 each correspond to new and aged junction wafers, with annealed junctions labeled by subscript L. Day 0 corresponds to the day of laser-annealing. }
\label{fig:fig4}
\end{figure*}
In this section we investigate the response of $R_\textrm{N}$ to lasing parameters at the wafer-scale in order to understand the laser-annealing mechanism. This is enabled by the automated JJ image recognition of our setup, which positions and focuses the JJ with respect to the beam within \SI{20}{\second} \cite{Durham_2018}. We utilize JJ test wafers with 3000 junctions similar to  Ref.~\cite{Kreikebaum_2020}. Across multiple wafers, we study four lasing parameters: power, spot displacement from the JJ, exposure time, and exposure repetition, as well as stability to aging (see Fig.~\ref{fig:fig4}, Fig.~\ref{fig:supp_fig2}, and Table.~\ref{table::table1}). Table.~\ref{table::table2} in the S.M. provides all parameters of each study.\par 
Based on previous studies of JJ thermal annealing, we hypothesize that laser-annealing locally heats the JJ and thickens the tunneling barrier, thereby increasing $R_\textrm{N}$ \cite{Koppinen_2007, Granata_2013, Granata_2007}. We first verify local heating by measuring the normalized resistance change ($\Delta R/R_{0}$) of JJs with three different areas with respect to lasing power (Fig.~\ref{fig:fig4}(a)). The normalized resistance change follows an exponentially plateauing function (red dashed line) that caps at $\Delta R/R_{0} = 1.8\%$. This trend is similar to that of low temperature ($<\SI{150}{\celsius}$) thermal annealing of JJs demonstrated by \cite{Koppinen_2007, Migacz_2003, Shiota_1992, Vettoliere_2020}. In order to correlate JJ temperature to laser power, we simulate the temperature ($T$) of a JJ directly illuminated by a Gaussian beam with waist \SI{0.81}{\micro\meter} of varying power ($P$) using COMSOL Multiphysics. We observe a linear increase $T(P)=2.47P+\SI{20}{\celsius}$ that reaches $\sim$\,$\SI{120}{\celsius}$ at $P=\SI{40}{\milli \watt}$. The resistance change at this temperature is similar to $\Delta R/R_{0}$ observed in the literature ~\cite{Migacz_2003, Shiota_1992}. Extending the comparison with thermal annealing, for lasing powers $>\SI{50}{\milli\watt}$ for our setup, which corresponds to JJ temperatures $>\SI{150}{\celsius}$ as given by $T(P)$, we expect a rapid increase in $R_\textrm{N}$. This is because in this temperature regime, accelerated growth of $R_\textrm{N}$ has been observed \cite{Vettoliere_2020, Migacz_2003, Shiota_1992}. \par
Next we investigate the heat absorption mechanism by studying the normalized resistance change with respect to laser spot displacement ($D$) from the JJ. The displacement is measured from below the junction center, as shown in the inset of Fig.~\ref{fig:fig4}(b). We observe that the measured $\Delta R/R_{0}$ (blue points) is maximized at a displacement of \SI{4}{\micro \meter}, which corresponds to the extension length of the Al electrodes beneath the JJs. This is due to two competing effects: increased reflection from Al/Al-O\textsubscript{x} as displacement is reduced and decreased heat transfer from the Si substrate as displacement is increased. We model the power loss from reflection by calculating the absorbed power with respect to beam displacement using Gaussian beam integration. We then multiply this absorption function with an exponentially decaying function ($H(D)=A\exp(-D/D_0)+B$) that models heat transfer, where $D_0$ is the characteristic decay length for thermal conduction (see Fig.~\ref{fig:supp_fig2}) \cite{Green_1995, Wang_2015}. We use this product function to fit the data (red dashed line). The reflection is minimized at $D>\SI{4}{\micro \meter}$ and $D_0=\SI{9.5}{\micro \meter}$, resulting in a maximum fitted $\Delta R/R_{0}$ at $D=\SI{5}{\micro \meter}$. The kink at $D=\SI{4}{\micro\meter}$ is due to the increased absorption as the spot moves away from the Al electrode and onto the Si. It can also be seen that when the beam is placed more than \SI{30}{\micro\meter} away from the JJ, the change in resistance approaches that of unannealed JJs (gray dashed line). In other words, $R_\textrm{N}$ is unaffected by a beam displaced more than \SI{30}{\micro\meter}. This demonstrates locality of the laser heating on the sub-millimeter length scale.\par
Based on the measurements, we suggest that the laser beam locally heats the JJs through the Si substrate. Heat absorption has been proposed to thicken the JJ tunnel barrier in studies based on thermal annealing \cite{Koppinen_2007}. Therefore, we measure the barrier thickness using high resolution transmission electron microscopy and fit the area-normalized $R_\textrm{N}$ to the exponential of barrier thickness (see Fig.~\ref{fig:supp_fig4} and Table.~\ref{table::table3}) \cite{Kleinsasser_1995}. From the fit, we estimate that a 30\% change in $R_\textrm{N}$ can orginiate from a $\sim$\si{\angstrom} change in tunnel barrier thickness. However, due to the non-uniformity of the barrier (dispersion$\sim\SI{0.4}{\nano\meter})$, we are unable to detect the corresponding increase in thickness (\SI{0.04}{\nano\meter}) of a $<10\%$ change in $R_\textrm{N}$ caused by laser-annealing. This non-uniformity makes it unlikely for a simple barrier thickening model to fully explain the microscopic mechanism. Instead, consideration of other microscopic factors, such as barrier height and chemical composition changes at the Al/Al-O\textsubscript{x} interface, is needed \cite{Granata_2007}.\par  
Lastly, we study how robust laser-annealing is with respect to aging. JJ aging refers to the increase in $R_\textrm{N}$ with exposure to air in time \cite{Schafer_1991, Koppinen_2007}. While the drift in $R_\textrm{N}$ due to aging is currently unavoidable, it is important that the resistance difference between laser-annealed and unannealed JJs is conserved for an extended period of time. For superconducting qubits, this translates to maintaining frequency differences between different qubits, which is important for frequency allocation. We study the robustness against aging as follows. We prepare two wafers, one with newly fabricated JJs (Wafer1) and the other with 130 day aged JJs (Wafer2). Wafer2 serves to show the drift in $R_\textrm{N}$ when aging effects are minimal. For each wafer, we probe the resistance of unannealed and laser-annealed JJs for a period of 30 days, stored in atmosphere. All four data groups are fit to an exponentially plateauing aging function, with the fit parameters given in Table.~\ref{table::table1} \cite{Kreikebaum_2020}. As can be seen in Fig.~\ref{fig:fig4}(c), aging effects are pronounced for Wafer1 (16\%) in comparison to Wafer2 (7\%). Furthermore, the standard deviation increases with respect to time, implying varying degrees of aging even amongst nominally identical JJs. However, on average, the difference in resistance change between unannealed and laser-annealed JJs for each wafer is maintained even after 30 days of aging. This demonstrates that laser-annealing is robust against aging. The wafer-scale study we have conducted in this section constitutes a first step in both wafer-scale applicability of laser-annealing, as well as investigating the underlying physical mechanism.\par  
%
%
%
%
\begin{table}[h!]
\begin{center}

    \begin{tabular}{| l | l | l | l |}
    \hline
     Sample	 	   			& \thead{Final\\Resistance\\Change} 		   		& \thead{Initial\\Resistance\\Change} 	 	   			&\thead{Aging Constant}\\ \hline\hline
	
    Wafer1\textsubscript{L}			& $21\pm 1.2\,\%$ 		& $9\pm 1.7\,\%$			& $10.40\pm 2.5\,$ days  \\ \hline
        Wafer1			& $16\pm 1.1\,\%$ 		& $3\pm 1.6\,\%$			&$8.72\pm 1.7\,$ days \\ \hline
        Wafer2\textsubscript{L}			& $11\pm 5.4\,\%$ 		& $3\pm 7.6\,\%$			& $41.15\pm 39.4\,$ days \\ \hline
        Wafer2			& $7\pm 4.0\,\%$ 		& $0\pm 5.6\,\%$			& $27.95\pm 21.5\,$ days \\ \hline

    \hline
    \end{tabular}

\end{center}
\caption{Fit parameters for Fig.~\ref{fig:fig4}(c). The fit is done using an exponential plateau function $\Delta R/R_{0}=A-B\exp(-t/\tau)$, where A corresponds to the Final Resistance Change, (A-B) the Initial Resistance Change, and $\tau$ the Aging Constant. The large fitting errors for Wafer2 are due to fitting an exponential at the tail. }
\label{table::table1}
\end{table}
\section{Conclusion}
We have constructed an automated laser-annealing apparatus using conventional microscopy components and demonstrated reliable frequency tuning of fixed-frequency transmon qubits. The high coherence of our transmons is preserved after laser-annealing. We have further observed an instance of coherence increase after laser-annealing, and performed TLS spectroscopy to investigate the change in defect features. These methods should be further explored towards treating defective qubits on multiqubit quantum processors.\par  
Furthermore, we have scaled up laser-annealing and studied effects of lasing parameters at the wafer scale. With this, we have put forth a model of local heating through the Si substrate. Additional studies with a change or etching of substrate underneath the JJ can help verify this model \cite{Degnan_2022,Chu_2016, Place_2021}. We have also demonstrated that laser-annealing is robust against aging, which is important for superconducting qubits since the time between laser-annealing and qubit measurement is non-negligible. Further studies are needed to correlate normalized $R_\textrm{N}$ to JJ barrier thickness. This can be realized using different JJ geometries or different JJ materials. Efforts in this direction are necessary since a thorough understanding of each fabrication and treatment step is ultimately required as qubit coherence times are pushed higher into the millisecond regime.\par 
\section{Acknowledgments}
\begin{acknowledgments}
The authors thank B. Marinelli for useful discussions regarding TLS. This work was funded by the U.S. Department of Energy, Office of Science, Office of Basic Energy Sciences, Materials Sciences and Engineering Division under Contract No. DE-AC02-05-CH11231 “High-Coherence Multilayer Superconducting Structures for Large Scale Qubit Integration and Photonic Transduction program (QIS-LBNL)”. Comsol simulations were performed in the Molecular Graphics and Computation Facility at UC Berkeley, which is funded by the Kavli Institute and NIH S10OD023532. Focused ion beaming (FIB) for HRTEM was conducted at the Surface Analysis Lab at University of Utah by Brian Van Devener and Randy C. Polson.
\end{acknowledgments}
\bibliography{LaserAnnealing}
\cleardoublepage
\onecolumngrid
\newcommand{\beginsupplement}{
    \setcounter{figure}{0} 
  \setcounter{table}{0}
    \renewcommand{\thetable}{S\arabic{table}}  
  \renewcommand{\thefigure}{S\arabic{figure}}
  \renewcommand{\theHfigure}{Supplement.\thefigure}
  \renewcommand{\theHtable}{S\arabic{table}}
}
\section{Supplemental Materials}
\beginsupplement
\subsection{Laser-Annealing Automation}
Automation of laser-annealing is based on JJ image recognition and auto-focusing. For a wafer of 3000 JJs spaced on a grid, we record the coordinates of 10 JJs and perform an affine coordinate transformation to obtain the coordinates of the other JJs. Each JJ is auto-focused by evaluating the image sharpness, which is calculated using the pixel width of macroscopic features such as the Al electrode arms. Afterwards, the JJ is centered to the laser using cross detection of Canny edges of the junction image. The centered, focused JJ is laser-annealed with the input parameters. Images of the JJs are captured for post-processing to exclude those that are improperly focused or centered. This process is also automated using an image structural similarity function that compares any given image with the image of a properly laser-annealed JJ. Images with a structural similarity index less than 0.97 correspond to poorly annealed JJs, and are hence excluded from the data set.\par
\subsection{Device Parameters and Fabrication}
Microwave properties (eigenmode, linewidth, coupling) of the bus, ROs, and qubits are simulated using Ansys HFSS. For the device shown in Fig.~\ref{fig:fig1}(b), qubit-resonator coupling g is \SI{50}{\mega \hertz} and the resonator line-width $\kappa$ is between 50 and \SI{200}{\kilo \hertz}. For the device used for TLS spectroscopy, g is \SI{80}{\mega \hertz} and the $\kappa$ is \SI{2.5}{\mega \hertz}.\par
The devices are fabricated on a Si substrate of resistivity $>\SI{10}{\kilo \ohm}$. After surface cleaning of the Si using piranha and buffered-oxide etch (B.O.E.), a \SI{200}{\nano \meter} layer of Nb is sputtered, on which the bus, ROs, and qubit capacitors are defined by electron-beam lithography and reactive ion etching. After an additional B.O.E. cleaning, Al/Al-O\textsubscript{x}/Al JJs with nominal critical current densities of \SI{500}{\nano\ampere\per\micro\meter\squared} are evaporated in the Manhattan style to form the transmon qubit. Galvanic contact between the Nb capacitor and the JJ is done by a bandaid process \cite{Potts_2001}. The diced device chips are cleaned using N-Methylpyrrolidone and then wirebonded to a copper cryopackage. A detailed process is given in Ref.~\cite{Blok_2021}. For the JJ test wafers, we remove the piranha cleaning and Nb deposition steps. These test wafers undergo a single deposition step of Al. Thus the JJs are shunted to large Al paddles which are used for resistance probing. A detailed process is given in Ref.~\cite{Kreikebaum_2020}.\par
\subsection{TLS Spectroscopy Cooldown Variations}
\begin{figure*}[h]
\includegraphics[width=\columnwidth]{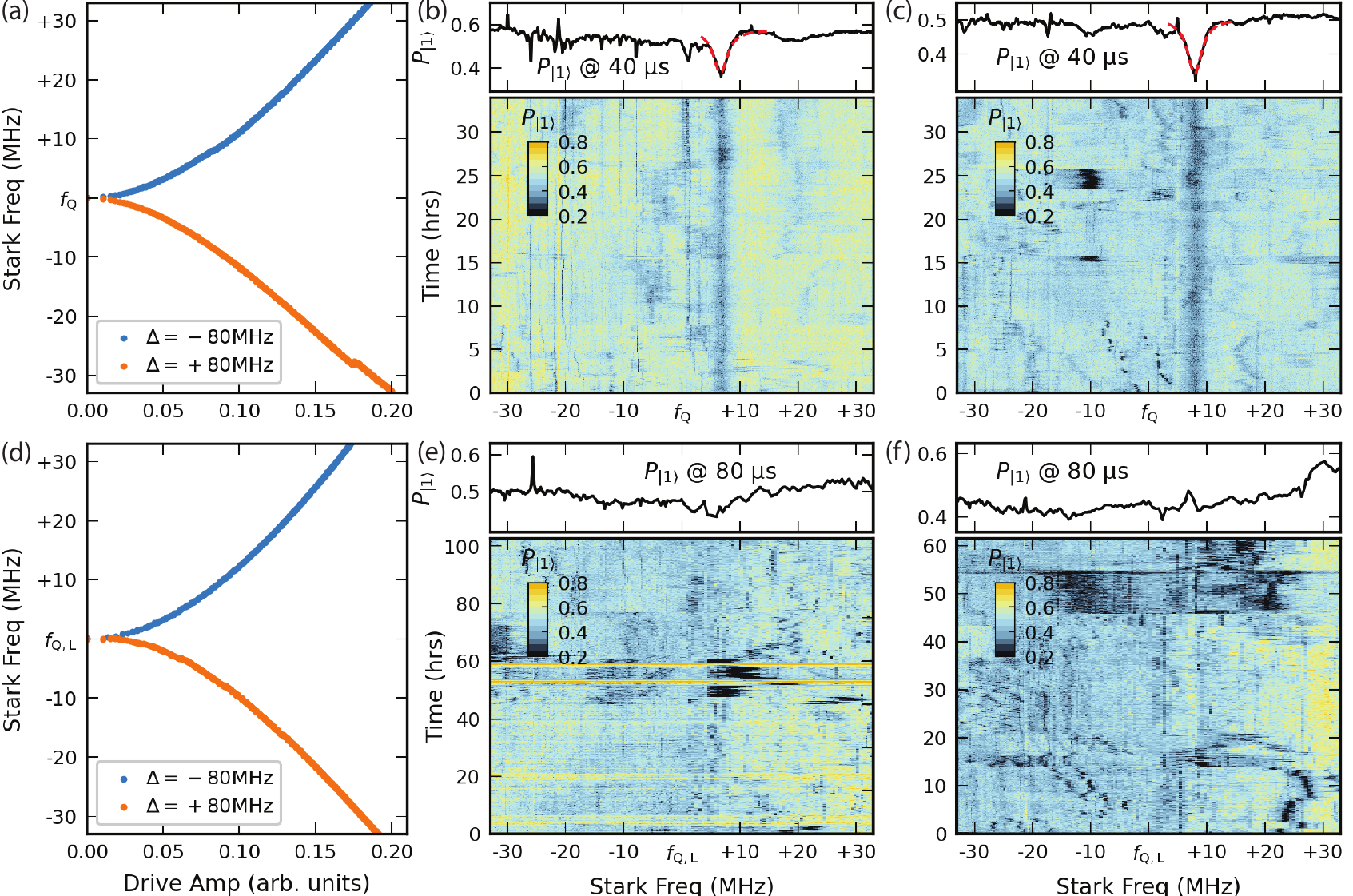}
\caption{(a) Calibration of AC Stark shift of transmon before laser-annealing. $\Delta=\pm\SI{80}{\mega\hertz}$ refers to Stark tone positively/negatively detuned from qubit: $f_\textrm{stark}=f_\textrm{Q}\pm\SI{80}{\mega\hertz}$. (b) TLS spectroscopy after thermal cycling of Fig.~\ref{fig:fig3}(b). The TLS feature lies at $f_{TLS}=f_Q+\SI{6.68}{\mega\hertz}$. (c) TLS spectroscopy after second thermal cycling. $f_{TLS}=f_Q+\SI{7.89}{\mega\hertz}$. (d) Calibration after laser-annealing. (e) TLS spectroscopy after thermal cycling of Fig.~\ref{fig:fig3}(c). No persistent TLS features are seen. (f) TLS spectroscopy after second thermal cycling. }
\label{fig:supp_fig1}
\end{figure*}

We calibrate the AC Stark shift by measuring the detuned frequency using a Ramsey sequence at each Stark tone amplitude. The Stark shift follows the frequency shift of an off-resonant Rabi drive in the driving frame: $\sqrt{\Omega^2+\Delta^2}$, where $\Omega$ is the drive amplitude and $\Delta$ is the detuning of the drive. Using $\Delta=\pm\SI{80}{\mega\hertz}$, we fit our calibration to the function: $\pm(\sqrt{(A\Omega')^2+\Delta^2}-\Delta)$, where $A$ is a fitted conversion parameter for the drive amplitude $\Omega'$. The calibration of the Stark shift before laser-annealing the transmon is shown in Fig.~\ref{fig:supp_fig1}(a). $A=\SI{432}{\mega\hertz}$ for $\Delta=-\SI{80}{\mega\hertz}$ and $A=\SI{416}{\mega\hertz}$ for $\Delta=+\SI{80}{\mega\hertz}$. We calibrate each time before conducting TLS spectroscopy. Here, we measure $P_{\ket{1}}$ at \SI{40}{\micro\second} which is around the $T_1$ of this transmon. The thermal cyclings of Fig.~\ref{fig:fig3}(a) are shown in Figs.~\ref{fig:supp_fig1}(b) and (c). The TLS spectroscopy was performed in the following order: Figs.~\ref{fig:supp_fig1}(b), Figs.~\ref{fig:supp_fig1}(c), and then Fig.~\ref{fig:fig3}(a). The consistent TLS mentioned above is seen across all three thermal cycles, with a frequency fluctuation of less than \SI{1}{\mega\hertz}. \par
We calibrate the Stark shift after laser-annealing the transmon. $A=\SI{459}{\mega\hertz}$ for $\Delta=-\SI{80}{\mega\hertz}$ and $A=\SI{416}{\mega\hertz}$ for $\Delta=+\SI{80}{\mega\hertz}$, shown in Fig.~\ref{fig:supp_fig1}(d). We measure $P_{\ket{1}}$ at \SI{80}{\micro\second} which is around the $T_1$ after laser-annealing. The thermal cyclings of Fig.~\ref{fig:fig3}(b) are shown in Figs.~\ref{fig:supp_fig1}(e) and (f). The TLS spectroscopy after laser-annealing was performed in the following order: Fig.~\ref{fig:fig3}(b), Figs.~\ref{fig:supp_fig1}(e), and then Figs.~\ref{fig:supp_fig1}(f). No persistent TLS features are seen across all three thermal cycles. For Fig.~\ref{fig:supp_fig1}(e), the horizontal orange lines at $50$ and  $\SI{60}{hours}$ are acquisition errors since $P_{\ket{1}}=1$ at all frequencies. We observe telegraphic TLS described in Ref.~\cite{Klimov_2018} in Fig.~\ref{fig:supp_fig1}(f) at $+\SI{20}{\mega\hertz}$ from $0$ to $\SI{10}{hours}$.\par
%
\subsection{Wafer Scale Characterization}
For the COMSOL simulation, we utilize a \textit{Heat Transfer in Solids} model with convection cooling and a Gaussian beam heat source to thermally simulate laser-annealing. 37.4\% of power reflected by silicon is taken into account. Convection cooling with room temperature of 20\textdegree C is input into the simulation. The beam is centered on the junction of interest, while another junction is placed 1.2 mm away to observe the temperature of neighboring junctions. The simulated temperature of the neighboring junction is \SI{22.9}{\celsius}. We fit the simulated temperature to the resistance data given in Fig.~\ref{fig:fig4}(a) using the function: $\Delta R/R_0=m-b\exp{(-T/T_0)}$, where $m$ is the final resistance change, $b$ determines the initial resistance, and $T_0$ is the characteristic temperature (red dashed line of Fig.~\ref{fig:fig4}(a)).\par
\begin{figure*}[h]
\includegraphics[width=\columnwidth]{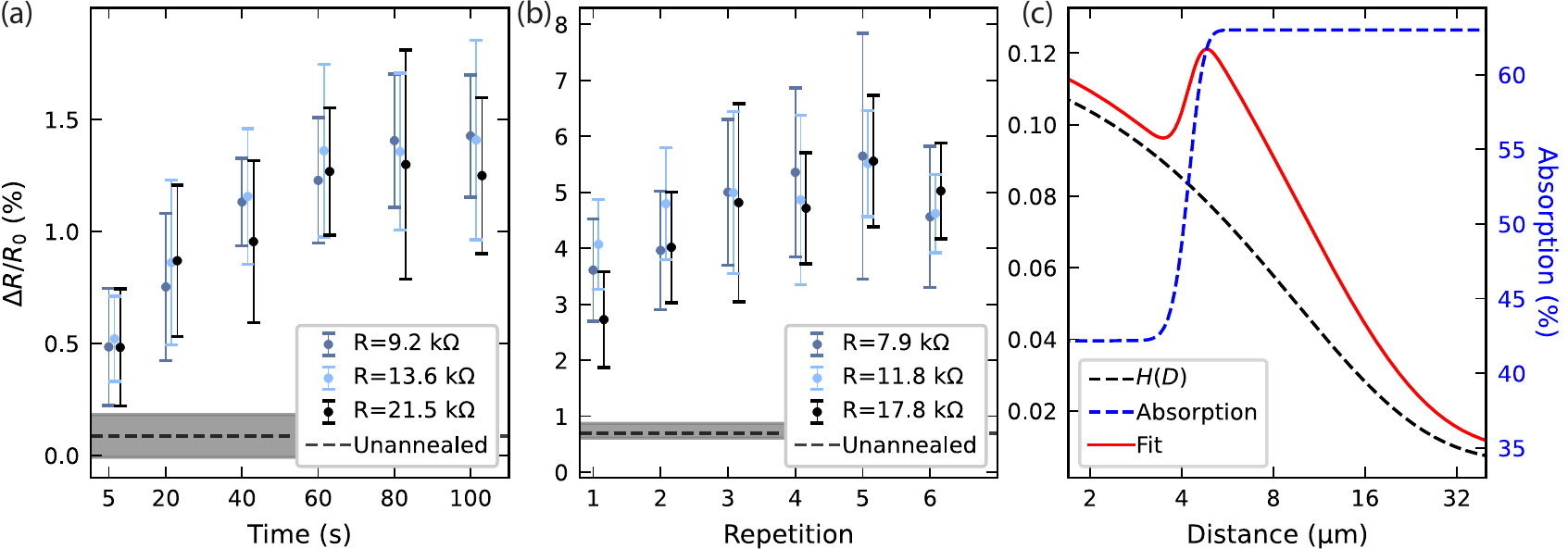}
\caption{(a) Percent JJ resistance change vs laser exposure time. (b) Percent JJ resistance change vs exposure repetition. (c) Calculated absorption and $H(D)$ used to produce the fit given in Fig.~\ref{fig:fig4}(b). The product is shown by the solid red line, which is identical to that shown in Fig.~\ref{fig:fig4}(b) but scaled down for clarity.}
\label{fig:supp_fig2}
\end{figure*}

The lasing parameters for each wafer-scale study are described in Table.~\ref{table::table2}. For each study, we vary a single parameter and keep all other parameters fixed. The normalized resistance change $\Delta R/R_0$ with respect to exposure time and repetition are shown in Figs.~\ref{fig:supp_fig2}(a) and (b). Both exhibit an exponentially plateauing increase, similar to Fig.~\ref{fig:fig4}(a). This suggests that the total amount of heat deposited on the JJ, which is a function of lasing power, exposure time, and exposure repetition, determines the increase in resistance. Each datapoint shown in Figs.~\ref{fig:fig4} and Fig.~\ref{fig:supp_fig2} exhibits uncertainty $\sim1\%$. This implies identically laser-annealed JJs will vary in resistance change. Therefore for precise tuning of $R_\textrm{N}$, JJs need to be iteratively annealed while $R_\textrm{N}$ is monitored. This will require multiple resistance probings of the qubit capacitor pads. The effects of this on qubit frequency and quality should be investigated.\par 
In Fig.~\ref{fig:supp_fig2}, the calculated absorption with respect to displacement (blue dashed line) as well as $H(D)$ (black dashed line) used for fitting Fig.~\ref{fig:fig4}(b) are shown. The absorption plateaus until $D=\SI{4}{\micro\meter}$, which corresponds to the displacement in which the beam moves away from the Al and onto the Si as mentioned in the main text. Due to this plateau, a kink arises when the absorption is multiplied with $H(D)$, as shown by the red line (equal to the fit shown in Fig.~\ref{fig:fig4}(b), but scaled down for clarity). Further analysis of the thermal conduction to the JJ ($H(D)$) with respect to beam displacement is needed.
\begin{table}[h]
\begin{tabular}{|l|l|l|l|l|l|l|}
\hline
Study        & Power                                                                  & Displacement                                                           & \thead{Exposure\\Time}                                                                   & \thead{Exposure\\Repetition}                                                             & \thead{Aging\\(Wafer1)}                                                             & \thead{Aging\\(Wafer2)}                                                             \\ \hline\hline
Applied Power        & Varied                                                                 & \SI{40}{\milli \watt} & \SI{40}{\milli \watt} & \SI{40}{\milli \watt} & \SI{40}{\milli \watt} & \SI{40}{\milli \watt} \\ \hline
Applied Displacement & \SI{0}{\micro \meter} & Varied                                                                 & \SI{0}{\micro \meter} & \SI{0}{\micro \meter} & \SI{0}{\micro \meter} & \SI{0}{\micro \meter} \\ \hline
Applied Exposure Time         & \SI{60}{\second}                     & \SI{60}{\second}                     & Varied                                                                 & \SI{60}{\second}                     & \SI{60}{\second}                     & \SI{60}{\second}                     \\ \hline
Applied Exposure Repetition   & 1                                                                      & 1                                                                      & 1                                                                      & Varied                                                                 & 1                                                                      & 1                                                                      \\ \hline
Utilized Wafer Age    & \SI{69}{\day}                        & \SI{88}{\day}                        & \SI{75}{\day}                        & \SI{71}{\day}                        & \SIrange{0}{30}{\day}              & \SIrange{135}{170}{\day}  \\ \hline        

\end{tabular}
\caption{Lasing parameters for each wafer-scale study. }
\label{table::table2}
\end{table}
\subsection{Microscopic Imaging}
\begin{figure}[h]
\includegraphics[width=.5\columnwidth]{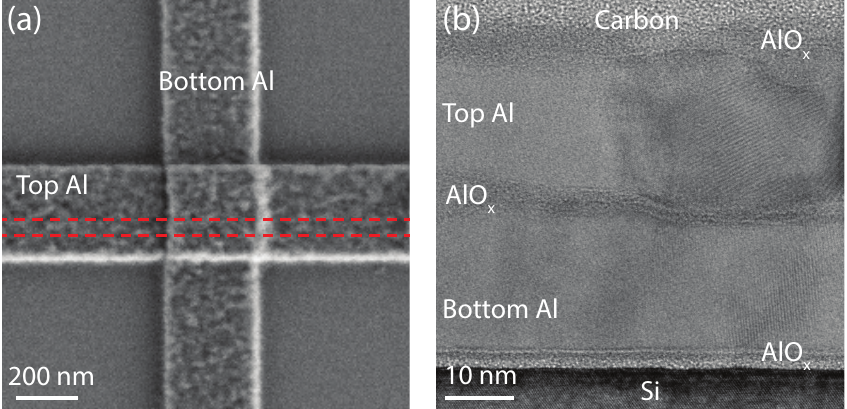}
\caption{(a) Scanning electron microscopy (SEM) of laser-annealed JJ. The red dash corresponds to the area on which a focused ion beam (FIB) is applied to acquire the cross-section of the JJ for high resolution  transmission electron microscopy (HRTEM). (b) HRTEM of JJ shown in (a). Each layer is given a corresponding label. The top layer of carbon is due to the FIB process. SEM and HRTEM images for the other four JJs given in Table.~\ref{table::table3} appear similar to that given here.}
\label{fig:supp_fig4}
\end{figure}

In order to correlate barrier thickness to resistance, we measure the surface area and barrier thickness of unannealed, laser-annealed, and thermally annealed JJs using scanning electron microscopy (SEM) and high resolution transmission electron microscopy (HRTEM). The SEM is a Zeiss Gemini Ultra SEM and the HRTEM is performed using the Transmission Electron Aberration-corrected Microscope 1 (TEAM1) at the National Center for Electron Microscopy (NCEM). The TEAM1 has a resolution of \SI{0.1}{\nano\meter}. For barrier thickness measurements, we utilize electron energy loss spectroscopy (EELS) using a Gatan Tridiem EELS spectrometer on the HRTEM results. The EELS measurements provide the JJ barrier thickness dispersion discussed in the main text. We utilize JJs on a 1x1 cm\textsuperscript{2} diced section of our JJ test wafers. The SEM and HRTEM images are shown in Fig.~\ref{fig:supp_fig4}. The measured $R_\textrm{N}$, area, and barrier thickness are given in Table.~\ref{table::table3}. \par
We fit the area-normalized $R_\textrm{N}$ to $\exp(t/\tau)$, where $t$ is the measured barrier thickness and $\tau$ is the characteristic barrier thickness. We obtain $\tau=0.39\pm\SI{0.23}{\nano\meter}$, which implies an increase of \SI{1}{\angstrom} in $t$ will increase $R_\textrm{N}$ by $\sim30\%$ as stated in the main text. However, the fit is limited due to the lack of statistics and large dispersion. This dispersion is due to the curvature of the tunneling barrier in our HRTEMs. Additional statistics are needed to properly correlate JJ barrier thickness to $R_\textrm{N}$.
%
%
%
\begin{table}[h]
\begin{center}
    \begin{tabular}{| l | l | l | l |}
    \hline
     Sample	 	   			& Resistance 		   		& Area 	 	   			&\thead{Barrier\\Thickness}\\ \hline\hline
	
    Unannealed			& \SI{7781}{\ohm} 		& \SI{0.0997}{\micro \meter \squared}			& 2.43 $\pm$ \SI{0.71}{\nano \meter} \\ \hline
        Unannealed			& \SI{5249}{\ohm} 		& \SI{0.1679}{\micro \meter \squared}			& 2.32 $\pm$ \SI{0.39}{\nano \meter} \\ \hline
        Laser-Annealed			& \SI{5979}{\ohm} 		& \SI{0.1125}{\micro \meter \squared}			& 2.44 $\pm$ \SI{0.54}{\nano \meter} \\ \hline
        \SI{400}{\celsius}-Annealed			& \SI{13735}{\ohm} 		& \SI{0.1967}{\micro \meter \squared}			& 2.44 $\pm$ \SI{0.23}{\nano \meter} \\ \hline
        \SI{400}{\celsius}-Annealed			& \SI{13867}{\ohm} 		& \SI{0.1835}{\micro \meter \squared}			& 2.64 $\pm$ \SI{0.07}{\nano \meter} \\ \hline

    \hline
    \end{tabular}
\end{center}
\caption{Measured resistance, area, and barrier thickness of five JJs. Two are unannealed, one is laser-annealed, and two are thermally annealed at \SI{400}{\celsius}.}
\label{table::table3}
\end{table}
\end{document}